# A Maximum Subsurface Biomass on Mars from Untapped Free Energy: CO and H$_2$ as Potential Antibiosignatures

Steven F. Sholes[a], Joshua Krissansen-Totton[a], and David C. Catling[a]

[a]*Department of Earth and Space Sciences and Astrobiology Program, University of Washington, Box 351310, Seattle, WA 98195, USA.*

## ABSTRACT

Whether extant life exists in the martian subsurface is an open question. High concentrations of photochemically produced CO and H$_2$ in the otherwise oxidizing martian atmosphere represent untapped sources of biologically useful free energy. These out-of-equilibrium species diffuse into the regolith, so subsurface microbes could use them as a source of energy and carbon. Indeed, CO oxidation and methanogenesis are relatively simple and evolutionarily ancient metabolisms on Earth. Consequently, assuming CO- or H$_2$- consuming metabolisms would evolve on Mars, the persistence of CO and H$_2$ in the martian atmosphere set limits on subsurface metabolic activity. Here, we constrain such maximum subsurface metabolic activity on Mars using a 1-D photochemical model with a hypothetical global biological sink on atmospheric CO and H$_2$. We increase the biological sink until the model atmospheric composition diverges from observed abundances. We find maximum biological downward subsurface sinks of $1.5{\times}10^8$ molecules cm$^{-2}$ s$^{-1}$ for CO and $1.9{\times}10^8$ molecules cm$^{-2}$ s$^{-1}$ for H$_2$. These covert to a maximum metabolizing biomass of $\lesssim 10^{27}$ cells or $\leq 2{\times}10^{11}$ kg, equivalent to $\leq 10^{-4}$-$10^{-5}$ of Earth's biomass, depending on the terrestrial estimate. Diffusion calculations suggest this upper biomass limit applies to the top few kilometers of the martian crust in communication with the atmosphere at low to mid latitudes. This biomass limit is more robust than previous estimates because we test multiple possible chemoautotrophic ecosystems over a broad parameter space of tunable model variables using an updated photochemical model with precise atmospheric concentrations and uncertainties from *Curiosity*. Our results of sparse or absent life in the martian subsurface also demonstrate how the atmospheric redox pairs of CO-O$_2$ and H$_2$-O$_2$ may constitute antibiosignatures, which may be relevant to excluding life on exoplanets.

## 1. INTRODUCTION

An open question about Mars is whether it currently hosts life in its subsurface. The *Viking* landers of the 1970s have been the only direct life detection missions on Mars and the consensus is that their results were negative (Klein 1978; Klein 1998). This is consistent with the poor habitability of the surface environment. Aside from plausible ephemeral briny solutions (Martín-Torres *et al.* 2015; Ojha *et al.* 2015; Rennó *et al.* 2009; Toner *et al.* 2014), the surface is cold and lacks lasting liquid water (Rummel *et al.* 2014) to provide a solvent for life. Also, bombardment from harsh galactic cosmic ray (GCR), solar ultraviolet (UV), and solar energetic proton (SEP) radiation in the upper few meters is enough to deactivate cells over timescales of 1,000-200,000 years depending on radioresistance and regolith materials (Cockell *et al.* 2000; Dartnell *et al.* 2007; Pavlov *et al.* 2012; Teodoro *et al.* 2018). Furthermore, highly-oxidizing perchlorate in the soil can become microbicidal under martian conditions (Wadsworth and Cockell 2017).

However, the possibility of extant martian life remains. Reported detections of transient methane (Formisano *et al.* 2004; Krasnopolsky *et al.* 2004; Mumma *et al.* 2009; Webster *et al.* 2015; Webster *et al.* 2018) have resulted in a surge of interest in the possibility of biological methanogenesis on Mars (Mickol and Kral 2016; Yung *et al.* 2018); although see Zahnle (2015; *et al.* 2011) for a dissenting view on the detections. Terrestrial microbes have also been found to use perchlorate oxidation-based metabolisms (Myers and King 2017). Subsurface life could also use deep subsurface $H_2$ gas sourced through cataclastic (McMahon *et al.* 2016), serpentinizing (Ehlmann *et al.* 2010), or radiolytic (Onstott *et al.* 2006) reactions as a potential energy source before reaching the surface but there is generally no way to constrain the extent of such life based on observables.

Instead, one way to constrain extant subsurface martian life is to assume that is in contact with the atmosphere through the porous regolith and to consider atmospheric disequilibrium. Atmospheric chemical disequilibrium is often discussed as a life-detection method for planetary atmospheres. For example, the co-existence of oxygen and methane in Earth's atmosphere is a biosignature because this out-of-equilibrium redox pair would not persist without replenishing fluxes of methane and oxygen (Hitchcock and Lovelock 1967; Krissansen-Totton *et al.* 2016; Lovelock 1965). If abiotic sources of $O_2$ and $CH_4$ can be ruled out, then the $O_2$-$CH_4$ pair is a disequilibrium biosignature. However, when the source of atmospheric disequilibrium is known to be abiotic, such as from photochemical or geological fluxes, and the predicted abiotic concentrations of the disequilibrium pair match that observed, then it suggests that life is not exploiting the atmospheric free energy.

Historically, it was argued that the martian atmosphere is essentially at equilibrium and therefore Mars is unlikely to support life (Lovelock 1988; Lovelock 1975; Lovelock 1979). However, with modern observations, it is understood that Mars' atmosphere has the largest thermodynamic disequilibrium in the Solar System aside from Earth, with an available free energy of ~136 J per mole of atmosphere (Krissansen-Totton *et al.* 2016). This free energy is predominantly attributable to the CO-$O_2$ redox pair produced abiotically by the photolysis of $CO_2$ and $H_2O$ in a thin, dry atmosphere.

There are compelling reasons to believe that martian life – even if it possesses radically different biochemistry to Earth life – would evolve to exploit the disequilibrium available from the redox pairs in Mars' atmosphere. The net reaction for the anaerobic CO metabolism is $CO + H_2O \rightarrow CO_2 + H_2$ (Techtmann *et al.* 2009) and is a relatively simple metabolism that only requires water, given a source of CO. The enzymes that catalyze this reaction on Earth, carbon monoxide dehydrogenases (CODHs), possess a variety of simple Ni-Fe or Mo active sites suggesting that they have evolved independently multiple times (Ragsdale 2004; Techtmann *et al.* 2009). Additionally, the genes encoding CODH are prolific with some 6% of sequenced microbial genomes having one or more copies of the Ni-Fe CODH. This has led some to argue that CO oxidation to be an evolutionarily ancient process, serving as both an energy and carbon source for the earliest forms of life in volcanic settings (Ferry and House 2005; Huber and Wächtershäuser 1997; Lessner *et al.* 2006; Ragsdale 2004). Indeed, genomic evidence suggests the last universal common ancestor (LUCA) possessed genes encoding CODH (Weiss *et al.* 2016). Additionally, there are modern methanogenic archaea that use CO for their entire metabolisms (Rother and Metcalf 2004).

Aerobic CO-metabolisms also exist, which oxidize CO through the net reaction $2CO + O_2 \rightarrow 2CO_2$ (Meyer and Schlegel 1983; Ragsdale 2004). These aerobic carboxydotrophs also use CODH enzymes that possess a variety of Mo-, Fe-, and Cu- activation sites (Jeoung and Dobbek 2007). Given the diversity of transition metal catalysts and the abundance of CO in the martian atmosphere, it is reasonable to expect that microbial life in contact in the atmosphere would evolve to exploit this available free energy.

A much smaller portion of Mars' atmospheric thermodynamic disequilibrium, due to the small concentration of $H_2$, is attributable to the $H_2$-$O_2$ redox pair. Microbial methanogenesis, oxidizing $H_2$ with $CO_2$, may be one of the most primitive metabolisms on Earth (Ueno *et al.* 2006; Weiss *et al.* 2016; Wolfe and Fournier 2018). Thus $H_2$-oxidation could also be a source of free energy for martian life, although it should be noted that CO provides >98% of the available atmospheric free energy.

All life as we know it, without exception, uses redox chemical reactions to generate energy (e.g. Falkowski *et al.* 2008). Given abundant free energy in the martian atmosphere, microbial life – if it exists – ought to take advantage of this 'free lunch' (Zahnle *et al.* 2011). The thermodynamic drive to equilibrium by chemosynthetic life suggests that certain molecules in abundance, such as the $CO$-$O_2$ and $H_2$-$CO_2$ redox pairs, constitute antibiosignatures (Catling *et al.* 2018; Wang *et al.* 2016) - evidence that actively metabolizing life is absent or sufficiently meager that it does little to perturb the atmosphere beyond its abiotic photochemical state. The persistence of this photochemically induced disequilibrium and the reasonable assumption that martian life would evolve a simple CO- or $H_2$-metabolism to take advantage of such a disequilibrium, suggests that any microbial life in communication with the atmosphere has not exploited this free lunch and thus it represents an antibiosignature.

Disequilibrium in Mars' atmosphere has previously been used to constrain a maximum extant biomass in the subsurface using a modified photochemical model to include biogenic sinks on CO and $H_2$ (Weiss *et al.* 2000). However, this study was potentially inaccurate because it used fixed values for photochemical model parameters tuned to modern abiotic martian conditions whereas one should use values that account for their large uncertainties. Instead, given the uncertainties on these parameters (described in Sec. 2.2), they could be tuned differently, but within allowable error bars, to an assumed Mars with biogenic sinks on CO or $H_2$. This procedure could potentially alter the outcome for calculated subsurface biomass.

Here we consider how much biomass could exist in the current subsurface and be feeding off atmospheric redox couples. A subsurface environment is assumed because persistent liquid water on Mars today can only exist in the subsurface where the geothermal gradient allows an aquifer (e.g. Dundas *et al.* 2014; Feldman *et al.* 2004; Grimm and Painter 2009; Mellon *et al.* 1997).

We use a novel approach by optimizing the photochemical model over a broad range of tunable parameter space that assumes a biosphere for calculating unknown variables. In addition to the significance of the optimization procedure, we additionally improve and expand upon the work done by Weiss *et al.* (2000) in four major ways: 1) We include sources and sinks for all metabolic reactants and products as opposed to only sinks on CO and $H_2$. 2) We use updated, more precise present day atmospheric compositions collected by the *Curiosity* rover with an improved photochemical model. 3) We account for multiple combinations of the plausible chemoautotrophic ecosystems that use different net metabolic pathways and use improved estimates for cellular maintenance energy. 4) Because the delivery of substrate gases to subsurface life (and hence its biomass) might be restricted by downward diffusion through the regolith, we develop an improved gas diffusion model and test over a broad range of surface properties and crustal gradients. Thus, our comprehensive results provide rigorous constraints on the maximum biomass that could exist on Mars today that is in communication with the atmosphere.

## 2. METHODS

To investigate the effects of biological sinks on the martian atmosphere, we use a one-dimensional photochemical code originally developed for modeling the early Earth by Kasting (1979) but our nominal model has since been modified and validated for the modern martian atmosphere (Sholes *et al.* 2017; Smith *et al.* 2014; Zahnle *et al.* 2008). This model is built to include C-H-O-N-S (carbon, hydrogen, oxygen, nitrogen, and sulfur) chemistry and improves upon the Nair *et al.* (1994) model used by Weiss *et al.* (2000) in three major ways: 1) for redox conservation, hydrogen-escape is balanced via an abiotic tunable deposition velocity ($v_{dep}$) of reactive oxidizing species to the surface rather than arbitrary O escape at the top of the model (Lammer *et al.* 2003; Lillis *et al.* 2017; Zahnle *et al.* 2008); 2) the implementation of diffusion-limited hydrogen escape, both of which are justified by Zahnle *et al.* (2008) with reference to data and models and by measurements by the *Mars Atmosphere and Volatile EvolutioN* (MAVEN) mission (Jakosky *et al.*

2018) where the measured H escape rate of $0.8\text{-}7.6 \times 10^8$ atoms cm$^{-2}$ s$^{-1}$ encompasses the estimated diffusion-limited escape rate of $(3.3\pm1.1)\times10^8$ H atoms cm$^{-2}$ s$^{-1}$ within uncertainties (Catling and Kasting 2017, pg. 148); and 3) implementing a lower boundary condition where the $CO_2$ concentration is no longer fixed.

With the exception of the $CO_2$ treatment (described in more detail below), we follow the same parameters as the aforementioned models. Atmospheric surface pressure is set to 6.5 mbar, close to present day global-average levels (Haberle *et al.* 2008). The model uses uniformly spaced 1 km resolution grids up to 110 km. Ionospheric chemistry is not directly modeled, but downward fluxes at the top of the model of key photolytically produced ionospheric species are included, namely NO, N, and CO. Oxygen escapes out the top of the model at a rate of $10^7$ O atoms cm$^{-2}$ s$^{-1}$ following (Zahnle *et al.* 2008) and is of the same order as measurements made by MAVEN ($\sim 3\times10^7$ O atoms cm$^{-2}$ s$^{-1}$ Jakosky *et al.* (2018)). Vertical transport of species is dominated by eddy diffusion. The temperature profile is approximated by $T = T_s - 1.4h$ for the lower 50 km and isothermal above; $T_s$ is the surface temperature (detailed below) and $h$ is the height above the surface (in km). Water vapor is kept at a constant relative humidity in the lower atmosphere to produce the observed 9.5 pr-μm (Zahnle *et al.* 2008)

Previous versions of the photochemical model held $CO_2$ at a constant mixing ratio, assuming it to be replenished by indefinitely large surface (e.g. polar caps) and subsurface reservoirs (e.g. Zahnle *et al.* 2008). Here, we allow the $CO_2$ mixing ratio to vary, rather than be fixed at a set concentration. This allows for more accurate behavior when biology is included because some metabolisms release or draw down $CO_2$. In particular, CO and $O_2$ are replenished through the photolysis of $CO_2$, so even though they would be drawn down as sinks in different metabolisms, they are being artificially replenished by an unrealistic injection of $CO_2$ (up to the set value) at all grids and time steps if $CO_2$ were fixed numerically in the model.

Validation tests were successfully performed to ensure that the updated model, where the $CO_2$ concentrations can vary, matches the previously validated models in their atmospheric structure and composition. Additional tests were done to compare each modeled metabolism with and without a fixed $CO_2$ mixing ratio (for both the "fixed parameter" and "optimized parameter" models described below). No significant differences were found other than a decrease in the maximum allowable sink for the net aerobic carboxydotrophy metabolism. When $CO_2$ concentrations are fixed, the maximum allowable biogenic sink for the net aerobic carboxydotrophy metabolism is unrealistic as it approaches the column-integrated production limit of CO due to the artificial replenishment of CO and $O_2$ in the atmosphere. All work presented here reflects the improved model with $CO_2$ concentrations no longer fixed, thus more accurately modeling the effect of each metabolism.

The influence of possible subsurface biological metabolic activity is modeled by ground-level fluxes in the model. For each of the metabolic ecosystems simulated, the model is run with that metabolism's substrate gases removed directly from the atmosphere via a fixed downward surface flux while we simultaneously inject its metabolic products directly into the atmosphere at the lower boundary. The model is then run to steady state and ground-level mixing ratios are compared with the observed abundances measured for the modern martian atmosphere (see Table 1) (Franz *et al.* 2017; Krasnopolsky and Feldman 2001; Webster *et al.* 2015; Webster *et al.* 2018). This process is repeated as the surface fluxes are incrementally increased until the modeled atmosphere diverges from the current atmospheric compositions within a 2σ (95%) uncertainty. We assume that this divergence then sets the limit on the extent of a metabolizing subsurface biomass because its activity would not violate the known atmospheric composition.

### 2.1 Fixed Parameter Model

We use the aforementioned nominal Mars model, which has been tuned and validated against the modern martian atmosphere (Sholes *et al.* 2017; Smith *et al.* 2014; Zahnle *et al.* 2008). There are five possible net chemoautotrophic metabolisms that could feasibly be living off atmospheric energy. We express these metabolisms as both individual metabolic pathways and as combined metabolic ecosystems as the Gibbs free energy produced through either net reaction are identical. The metabolisms (Mets.) are:

• Met. 1: Anaerobic CO metabolism only:

$$CO + H_2O \rightarrow CO_2 + H_2 \qquad \textbf{1}$$

• Met. 2: Methanogenesis only:

$$CO_2 + 4H_2 \rightarrow CH_4 + 2H_2O \qquad \textbf{2}$$

• Met. 3: Anaerobic CO metabolism and methanogenesis (i.e., 4×Met.1 + Met. 2):

$$4CO + 2H_2O \rightarrow 3CO_2 + CH_4 \qquad \textbf{3}$$

• Met. 4: Aerobic carboxydotrophy or an equivalent net CO metabolism, methanogenesis, and methanotrophy ecosystem (i.e., 4×Met.1 + 2×Met.2 + methanotrophy (
$CH_4 + 2O_2 \rightarrow 2H_2O + CO_2$ )):

$$2CO + O_2 \rightarrow 2CO_2 \qquad \textbf{4}$$

• Met. 5: Aerobic hydrogenotrophy or an equivalent net methanogenesis and methanotrophy ecosystem (i.e., Met.2+methanotrophy):

$$O_2 + 2H_2 \rightarrow 2H_2O \qquad \textbf{5}$$

These end-member cases represent the full range of possible martian ecosystems subsisting off atmospheric free energy. Weiss *et al.* (2000) only modeled Met. 4 and Met. 5 as they believed the most energetic reactions would likely use $O_2$ as an oxidant. We do not consider a methanotrophy-only metabolic system (or combinations of ecosystems that include methanotrophs without methanogens) because it is infeasible given that the concentration of background methane is <1 ppb (Webster *et al.* 2015; Webster *et al.* 2018) and thus will produce a much smaller maximum biomass than other metabolisms. Reported transients of methane >10 ppb are localized and controversial in nature (Zahnle *et al.* 2011). Notably, in the *Curiosity* tunable diode laser results of Webster *et al.* (2018), such transients are only found sometimes in a direct measurement protocol but never found in a more sensitive enrichment protocol. Thus, these results demand skepticism as the transients appear to be more of a function of measurement protocol than of Mars. Similarly, we do not test metabolisms that produce species not observed in the current martian atmosphere (e.g. sulfate reducers producing $H_2S$).

*2.2 Optimized Parameter Model*

The methodology for what we call the "fixed parameter" model (Section 2.1) may not provide a rigorous maximum biogenic sink because it is tuned to a modern abiotic Mars with fixed unknown parameters (described below) that have been tuned in the model. If life is actively metabolizing on Mars, then the tuned parameter values could be different to what is commonly assumed, and their incorrect tuning could mask biogenic sources/sinks. Thus, the "fixed parameter" model underestimates the maximum biogenic sink.

We have identified three main tunable variables in the nominal model where the value has significant impact on the atmospheric composition and is either unknown or not agreed upon in the literature: surface temperature, ionospheric flux, and deposition velocity. Of these, mean surface temperature is the most constrained based on observed measurements, but *global* mean surface temperature is dependent on factors such as global circulation model integrations and atmospheric dust content (Haberle 2013). Ionospheric fluxes of the odd nitrogen species (NO and N) are dependent on their concentrations and ratio, which are not entirely known. Krasnopolsky (1993) considers cases where the upper boundary flux of NO into the neutral atmosphere is both nonexistent and appreciable ($10^7$ molecules cm$^{-2}$ s$^{-1}$), so we consider the range $10^1$-$10^8$ molecules cm$^{-2}$ s$^{-1}$ consistent with sensitivity testing by Smith *et al.* (2014). A flux of CO into the upper atmosphere is set to be equal to that of NO to conserve redox balance while N is set to 10% that of NO (Sholes *et al.* 2017; Zahnle *et al.* 2008).

Deposition velocity of reactant species is the dominant free parameter of the model. It is a constant used to simulate mixing, molecular diffusion, adsorption, and reactions of species with the surface regolith in a single variable. Higher $v_{dep}$ indicates a more chemically reactive species (Catling and Kasting 2017; Seinfeld and Pandis 2016). In a physical system, the deposition velocity for each species would be different, but these variations would be small so a single $v_{dep}$ is assumed for all reactive species. This is considered a good approximation (Zahnle *et al.* 2008) as $H_2O_2$ and $O_3$ are the primary reactants that must have the $v_{dep}$ tuned for each model run to produce the modern abundances for the primary atmospheric constituents and computational restrictions prevent optimization of $v_{dep}$ over 30+ species. $CO_2$, CO, $O_2$, and $H_2$ are considered non-reactive with the surface and their $v_{dep}$=0 cm s$^{-1}$ (See Zahnle *et al.* 2008 for more on $O_2$ dry deposition and Sholes *et al.* 2017 for a review on CO deposition). For each species, $v_{dep}$ is a fixed value, but is used to compute a variable surface flux ($\Phi$) for each chemically reactive species, which depends on species abundance as $\Phi = -v_{dep}\, n_{i,surf}$, where $n_{i,surf}$ is the surface number density for species $i$ (Catling and Kasting 2017; Seinfeld and Pandis 2016). This net oxidant sink balances the atmospheric redox budget by countering hydrogen escape to space, which oxidizes the atmosphere (Lammer *et al.* 2003; Seinfeld and Pandis 2016; Zahnle *et al.* 2008). The biological consumptions of CO, $O_2$, and $H_2$ are included using a separate fixed flux term.

To test whether optimizing the tunable parameters affects the maximum biological sink in the modern martian atmosphere, we perform a grid search and optimization procedure through the parameter space for the five net ecosystems described in Section 2.1. We incrementally increase the net metabolic source/sink fluxes as described for the nominal model, but at each increment we perform a grid search of model parameter space to find whether any possible sets of parameters (for temperature, ionospheric flux, and deposition velocity) can reproduce the observed martian atmosphere while incorporating the biogenic fluxes; what we describe as our "optimized parameter' model. At every grid point in parameter space we perform an optimization procedure (scipy.optimize.minimize in Python) to ensure that all regions in parameter space were explored and not missed by the coarse grid search. Table 2 summarizes the plausible parameter ranges, bounds, and number of grid points for each parameter in addition to the tuned values that produce the modern atmospheric concentrations.

The optimization procedure minimized the Chi-Squared ($\chi^2$) value comparing the modeled atmospheric composition with the measured atmospheric composition and uncertainties in Table 1. This procedure was continued until the modeled atmospheric compositions diverged from the observed composition, regardless of parameter values.

*2.3 Atmospheric Diffusion*
Once maximum permissible biogenic sinks are found in both the "fixed parameter" and "optimized parameter" versions of the model, we consider how these fluxes compare with passive atmospheric diffusion into the crust. This is because the dry, harshly irradiated, and constantly oxidized martian surface is ostensibly uninhabitable (see Sec. 1) so any extant life would need to take advantage of aquifers in the subsurface. There may be small pockets of briny fluids in the upper few kilometers (Orosei *et al.* 2018), but a deep water-saturated layer should exist within 10 km of the surface as a consequence of the geothermal gradient (Clifford 1993; Clifford *et al.* 2010; Michalski *et al.* 2013). Here, we calculate to what depth microbes could plausibly exploit the atmospheric free energy and thereby assign a maximum crustal depth to our upper biomass limit (i.e. the biomass of microbes living below this depth would be limited by diffusion of substrate gases and would therefore have smaller surface sinks than the calculated maximum fluxes that ignore a diffusion restriction).

To calculate the diffusion flux of CO and $H_2$ at depth (the potential flux), we use a modified form of Fick's laws which assume mean free paths of diffusing gases are greater than the typical pore size such that Knudsen diffusion dominates. In one dimension, the general form of Fick's second law states:

$$\frac{\partial n_i}{\partial t} = \frac{\partial}{\partial z}\left( D_i(z)\frac{\partial n_i}{\partial z} \right)$$  **6**

where $n_i$ is the number density for species $i$ (molecules cm$^{-3}$), $z$ is the depth (cm), $t$ is time (s), and $D_i$ is the diffusion coefficient (diffusivity, cm$^2$ s$^{-1}$). In steady state, the concentrations will not change with time, so the left-hand side of this equation is set to 0. The solution for the concentration gradient, $\frac{\partial n_i}{\partial z}$, can then be used to find the potential flux, $F_i$, via Fick's first law:

$$F_i = -D_i(z)\frac{\partial n_i}{\partial z}$$  **7**

The physical environment over which the gas diffuses changes greatly through the upper ~10 km of martian regolith. This means that that the diffusivity, the controlling factor, changes drastically with increasing depth due to increasing pressure and temperature and can be written as:

$$D_i = \frac{\varepsilon(z)\,r(z)}{3\,\tau(z)}\sqrt{\frac{8\,R\,T(z)}{\pi\,m_i}}$$  **8**

where $\varepsilon(z)$, $r(z)$, $\tau(z)$, $T(z)$ are the porosity, pore size (cm), tortuosity (path twistiness), and temperature (K) at a given depth. $R$ is the gas constant and $m_i$ is the molar mass of the species. We follow Stevens $et\ al.$ (2015); (2017) in assuming that porosity and tortuosity decrease exponentially with depth with $\varepsilon(z) = \varepsilon_0\,e^{\frac{-z}{k}}$ and $\tau(z) = \tau_0\,e^{\frac{-z}{3k}}$ where $\varepsilon_0$ and $\tau_0$ are the surface porosity and tortuosity respectively and $k$ is a scaling factor set by the pore closure depth (Clifford 1993). Temperature is modeled as a linear geothermal gradient, $T(z) = a_T\,z + T_0$ where $T_0$ is the modeled surface temperature and $a_T$ is the gradient (Michalski $et\ al.$ 2013). We assume pore size follows a constant pressure gradient (Stevens $et\ al.$ 2015), $r(z) = a_r\,z + r_0$ where $a_r$ is the gradient defined as $\frac{r_0}{-k}$ such that pore size is 0 at the pore closure depth and $r_0$ is the average surface pore size. We perform these calculations over a range of plausible gradients and surface parameter values (detailed below).

If we consider a layer of biota sustaining off atmospheric energy, living at a depth $z_b$, and metabolizing all available reactants (CO or $H_2$), Eqn. 6 can be numerically solved as a boundary value problem. Using observed surface number densities, $n_i(0)$, and setting $n_i(z_b)=0$ we numerically solve for $n(z)$ (scipy.integrate.solve_bvp in Python) in steady state. We can then use Eqn. 7 to solve for potential flux as a function of biotic layer depth ($z_b$). In practice, microbes cannot metabolize with substrate conditions of zero within the biologically active layer, but observations of Antarctic soils have found that microbes can metabolize with CO concentrations of 20 ppb (Ji *et al.* 2017). Setting $n_i(z_b)$ to the equivalent of 20 ppb for CO under Antarctic conditions (approximated as 1 bar and 255 K) results in nearly identical potential fluxes as the $n_i(z_b)=0$ case demonstrating that a zero-concentration lower boundary condition is a reasonable assumption.

Given the uncertainties of the physical surface conditions and gradients, we employ a Monte Carlo simulation to test the potential flux under both a range of $z_b$ and physical parameter values (for porosity, tortuosity, average pore size, pore closure depth, and temperature). For each possible $z_b$, spaced every 10 m between 0-10 km, we perform 10,000 diffusion calculations where we assume each parameter has a uniform distribution bounded by the parameter ranges described above and sample these distributions randomly. From these outputs, we obtain a likelihood distribution for the diffusion flux as a function of biological layer depth.

Surface porosity ranges from $\varepsilon_0$=0.2-0.6, with low values characteristic of lunar estimates and high values approximating *Viking* lander observations (Clark *et al.* 1976; Clifford 1993; Sizemore and Mellon 2008). Estimated surface tortuosity ranges from $\tau_0$=1.5-2.5 as measured in Mars-regolith simulant soil (Sizemore and Mellon 2008). We take an average surface pore size range of $r_0$=$10^{-3}$-$10^{-4}$ cm based around the typically cited value of $6\times10^{-4}$ cm (Weiss *et al.* 2000) and experimental pore size distributions using glass beads (Sizemore and Mellon 2008). The pore closure depth due to compaction ranges from $k$=6-26 km depending on how saturated the crust is with water. The extremes are most likely over/under estimates and a value of 10 km is typically assumed (Hanna and Phillips 2005). We test temperature gradients of 10-30 K km$^{-1}$ (Michalski *et al.* 2013).

## 3. RESULTS

Figure 1 shows the effects of increasing the subsurface biological sinks for Mets. 1-3 while Figure 2 shows the results for Mets. 4 and 5. In all cases, as the sink is ramped up, the resulting modeled atmospheric composition eventually diverges from the $2\sigma$ uncertainty of the measured modern atmospheric composition. For example, in the fixed-parameter model Met. 1 case, the CO abundance falls below the observed $2\sigma$ abundance of 744 ppm at a downward biologic flux of $8.4\times10^5$ CO molecules cm$^2$ s$^{-1}$. In this case, CO is considered the "break species" as CO diverges from the observed concentrations at a lower biological sink than the other measured constituents ($O_2$, $H_2$, $CH_4$, $CO_2$). The flux at which this divergence occurs is the maximum permissible flux where the modeled metabolism can still replicate observed abundances and is listed in Table 3 for the fixed-parameter model Met. 1 case.

In this example (Met. 1, anaerobic CO metabolism, fixed-parameter model), as the biological sink on CO is ramped up, CO concentrations decline, $O_2$ abundances steadily rise, and $CO_2$ abundances eventually decline as well. These behaviors are the result of the simulated biological draw-down on CO. $CO_2$ is readily converted into CO and $O_2$ through photolysis, thus removal of CO through high biological sinks inhibits the recombination reactions back into $CO_2$ and the injection of $CO_2$ as a product of the metabolism eventually becomes insufficient to counteract the removal of CO. This leads to low CO, excess $O_2$ and eventually low concentrations of $CO_2$ (Catling and Kasting 2017 pgs. 338-340).

Except for metabolisms that produce $CH_4$, in all other metabolisms in the fixed-parameter model runs, CO concentrations deviate from observations at the lowest flux levels ($CH_4$ deviates at lower fluxes in metabolisms that produce it). This is due to either the direct biogenic sink on CO or the removal of $O_2$ (a byproduct of $CO_2$ photodissociation), which throttles the recombination of $CO_2$ and leads to excess CO. Because CO has the smallest observational uncertainty, the nature of the $\chi^2$-calculation in the optimized-parameter model will end up attempting to constrain CO more precisely than the other species. Thus, the optimized-parameter model runs generally have $CO_2$ concentrations that deviate at lower fluxes except for $CH_4$-producing metabolisms and the aerobic $H_2$-metabolism (where $O_2$ and $H_2$ are drawn down and eventually diverge from observations).

The maximum biogenic sink fluxes for each of the net ecosystems with both the fixed- and optimized-parameter models are summarized in Table 3. We find that the maximum downward flux on CO or $H_2$ ranges from $10^5$-$10^8$ molecules $cm^2$ $s^{-1}$. As expected, the optimized version of the model produced maximum fluxes that were consistently higher than their fixed-parameter model counterparts. The exceptions are Mets. 2 and 3 which had comparable maxima due to $CH_4$ building up too high and the tunable parameters unable to effectively remove $CH_4$. The anaerobic $H_2$-metabolism (Met. 5) allowed for the greatest downward biogenic sink of $1.9 \times 10^8$ $H_2$ molecules $cm^2$ $s^{-1}$ in the optimized-parameter model with the aerobic CO-metabolism (Met. 4) having a similar maximum of $1.5 \times 10^8$ CO molecules $cm^2$ $s^{-1}$.

Our fixed-parameter model results show a maximum downward sink of $2 \times 10^6$ CO molecules $cm^2$ $s^{-1}$ for Met. 4 and $4 \times 10^7$ $H_2$ molecules $cm^2$ $s^{-1}$ for Met. 5, which are, respectively, 30 times smaller and 2 times greater than the maximum sinks from Weiss *et al.* (2000) (which tested only these two metabolisms). While it is expected, given our model improvements, that we find an improved result for Met. 4, it may seem counterintuitive that we find a slightly greater sink for Met. 5 (methanogenesis and methanotrophs) compared with Weiss *et al.* (2000). This stems from our use of 2σ uncertainties on atmospheric abundances rather than an arbitrary 25% of the observations (the ratio of the measured uncertainty to the value) they used (at 1σ we find a maximum sink of $\sim 2 \times 10^6$ $H_2$ molecules $cm^2$ $s^{-1}$ for the fixed model Met. 5).

Maximum biogenic fluxes can be used to estimate maximum extant biomass. The conversion from flux to biomass is done using:

$$M_{max} = \frac{\Phi_{max} \, A_{Mars} \, \Delta G}{q_{BPR} \, N_A}$$

**9**

Here, the maximum metabolizing biomass ($M_{max}$) is equal to the maximum biogenic sink in model units ($\Phi_{max}$; molecules cm$^{-2}$ s$^{-1}$) times the surface area of Mars ($A_{Mars}$; cm$^2$) over the ratio of the basal power requirement (BPR) for life ($q_{BPR}$; kJ s$^{-1}$ cell$^{-1}$) to the specified reaction free energy ($\Delta G$; kJ mol$^{-1}$) for each metabolism times Avogadro's number ($N_A$; molecules mol$^{-1}$). The Gibbs free energies for each equation are -24 kJ mol$^{-1}$ CO, -24 kJ mol$^{-1}$ H$_2$, -48 kJ mol$^{-1}$ CO, -264 kJ mol$^{-1}$ CO, and -215 kJ mol$^{-1}$ H$_2$ for Equations 1-5 respectively (calculated with the freely available database model of Krissansen-Totton *et al.* 2016 and using the nominal surface conditions and gas concentrations for Mars).

Quantifying the minimal basal power requirement (BPR) of organisms is currently the subject of much debate because all the contributing factors are not well characterized (see Discussion) (Bradley *et al.* 2018; Hoehler and Jorgensen 2013; Kempes *et al.* 2017; LaRowe and Amend 2015; Lever *et al.* 2015). Many estimates have been made, but we elect to use the value of 3×10$^{-23}$ kJ s$^{-1}$ cell$^{-1}$ from Lever *et al.* (2015) and LaRowe and Amend (2015). This value is the smallest measured BPR from measurements of sulfate-reducing bacteria in anoxic marine sediments (Lever *et al.* 2015). However, the BPR could potentially be as low as 1×10$^{-24}$ kJ s$^{-1}$ cell$^{-1}$ which is a theoretical limit to prevent racemization of amino acids. BPRs less than this value may be able to sustain individual cells, but are characterized by population decay and thus not appropriate for characterizing a robust minimum biomass (one that is neither growing nor decaying).

Using this minimum BPR value, we find an upper limit of ~10$^{27}$ cells that could be supported of the available free energy of the martian atmosphere. This corresponds to the maximum biogenic sink for both the optimized-parameter Met. 4 (aerobic carboxydotrophs) and Met. 5 (aerobic hydrogenotrophs) models. These provide similar maximum biomass values as their slightly varying max fluxes and free energy converge on a similar maximum biomass based on Eqn. 6. Additional conversions from this estimated cellular biomass into a metric biomass are provided in the Discussion.

The likelihood functions for our Monte Carlo simulation on diffusion of CO and H$_2$ into the subsurface are plotted in Fig. 3. The depths at which these potential fluxes equate to our calculated maximum downward biogenic sinks (2.1×10$^8$ molecules cm$^{-2}$ s$^{-1}$ for CO-sink metabolisms and 1.3×10$^8$ molecules cm$^{-2}$ s$^{-1}$ for H$_2$) represent the maximum plausible depth of any microbes subsisting off atmospheric free energy. The intercept depths for the CO-sink and H$_2$-sink metabolisms with the median flux distribution are ~6.5 km and ~1 km respectively. Below these depths, the limiting factor for determining the maximum extant biomass is the flux of reactants rather than the photochemistry.

## 4. DISCUSSION

Our calculated maximum extant metabolizing martian biomass of ~$10^{27}$ cells is difficult to interpret in isolation and so we compare it to Earth's total biomass. Typical estimates for Earth's biomass are classically given in terms of petagrams of carbon mass (Pg C = $10^{15}$ g C) and range from 550 Pg C (Bar-On *et al.* 2018) to 720 Pg C (Kallmeyer *et al.* 2012). Converting these values into an estimated number of cells requires an assumption on the average cellular carbon weight. Average cellular carbon mass ranges from 5-85 fg C cell$^{-1}$ for microbial cells (Bakken 1985; Kallmeyer *et al.* 2012), so an estimated total number of cells on Earth ranges from ~$6.5 \times 10^{30}$ to ~$1.4 \times 10^{32}$ cells. Thus, our calculated maximum cellular biomass estimate of $10^{27}$ cells is 2-40$\times 10^{-5}$ of the Earth's biomass or roughly one hundred thousandth of the estimated total of Earth's biomass.

Converting this maximum cellular biomass for Mars into a metric biomass (in kg) requires an average microbial mass. Assuming an average dry mass of 14 fg cell$^{-1}$ (Hoehler and Jorgensen 2013; Kallmeyer *et al.* 2012), this amounts to ~$10^{11}$ kg total mass, assuming dry cell mass constitutes 20% of total cellular mass (Bratbak and Dundas 1984). In more tangible terms, this total mass is the equivalent to approximately 1 million blue whales (taking the average mass of a blue whale as $1.85 \times 10^{5}$ kg). While this amount of blue whales worth of microbial life on Mars appears large, it is vanishingly small compared to the total biomass on Earth today and is an upper limit based on generous assumptions that give the largest biomass.

As a comparison to the previous estimate, Weiss *et al.* (2000)'s work produced maximum biomass of ~$10^{10}$ kg or ~120,000 blue whale masses. However, this value uses their calculated energy flux at 10 m depth, assumes only 10% metabolic efficiency, and use a "typical maintenance energy" value of ~$10^{-4}$ kJ g$^{-1}$ s$^{-1}$ (~$10^{-18}$ kJ cell$^{-1}$ s$^{-1}$ using our assumed average cell mass). While it may be reasonable to assume such a power requirement and biological efficiency, this does not place an upper limit on the maximum possible biomass as microbes have been discovered subsisting off smaller energy fluxes. Indeed, if we take the maximum flux values used in their biomass calculations and apply our updated values for basal power requirement, full metabolic efficiency, and updated Gibbs free energy then their maximum biomass would be ~$10^{29}$ cells equivalent to ~$10^{13}$ kg or approximately 70 million blue whale masses.

The maximum biomass we calculate is likely an overestimate. Other considerations that we have not included in this methodology would lower the upper bound on biomass. It is assumed here that the metabolisms work at 100% efficiency - which is unlikely as energy is wasted in 'spillover' reactions and Darwinian evolution favors survival over efficiency (Weiss *et al.* (2000) assume the biological oxidation process works at ~10%). We also use basal power requirements to estimate biomass, but this is the minimum energy required to keep cells alive and does not account for additional energy costs such as reproduction. Abiological oxidation reactions with the regolith should contribute a small portion of our calculated maximum reactant sinks, thus the maximum biological sinks would be smaller than reported here.

There is also a great uncertainty in the minimum power requirements necessary to sustain actively metabolizing life. Older literature typically used the term "minimum maintenance energy" (including Weiss et al. 2000), which is defined as the minimum energy flux required to sustain a steady-state population without growth. However, recent work has contested this term as it includes both energy that is useful for the cells, e.g. motility and synthesis of bioimportant macromolecular compounds, along with energy that is energy that is wasted in "spill-over" reactions and thus does not accurately represent the actual minimal energy required for life to survive. Following Hoehler and Jorgensen (2013), we elected to use the basal power requirement, which is the energy flux required for the minimal amount of cellular functions to maintain a metabolically active cell.

Both the minimum maintenance energy and basal power requirement are notoriously difficult to measure. Lab studies have yielded maintenance energies of order $10^{-5}$ kJ $g^{-1}$ $s^{-1}$ (Tijhuis *et al.* 1993) to downwards of $10^{-7}$ kJ $g^{-1}$ $s^{-1}$ (Scholten and Conrad 2000). The upper end of these values are especially problematic as it is ~26 times greater than the maintenance energy of a human body, which is not operating near its limits (Hoehler 2004), and because it is temperature dependent. Minimum power requirements are especially difficult to characterize in natural settings (Kempes *et al.* 2017; Onstott *et al.* 2014; Van Bodegom 2007), and recent work has shown that the BPR is a function of cellular volume (Kempes *et al.* 2016). Measurements of sulfate-reducing bacteria in anoxic marine sediments show BPRs as low as $3 \times 10^{-23}$ kJ $s^{-1}$ $cell^{-1}$ (Lever *et al.* 2015). This lower value agrees well with other recorded lower bounds on BPR of ~$5 \times 10^{-23}$ kJ $s^{-1}$ $cell^{-1}$ (LaRowe and Amend 2015; Marschall *et al.* 2010). A back-of-the-envelope attempt by LaRowe and Amend (2015) at arriving to a theoretical lower limit provides a BPR of ~$1 \times 10^{-24}$ kJ $s^{-1}$ $cell^{-1}$ to prevent racemization of amino-acids and thus population decay. However, the authors do note that this may not be a true limit and could be off my orders of magnitude. Additionally, a BPR of ~$2 \times 10^{-22}$ kJ $s^{-1}$ $cell^{-1}$ is more characteristic of microbes in these ultra-low energy environments rather than the lowest cell-specific BPR of ~$3 \times 10^{-23}$ kJ $s^{-1}$ $cell^{-1}$. We have also been assuming in our calculations that the microbes are living at their lower limits with no population growth with very long turnover rates, which may very well be not the case.

It is also important to note that this is a rigorous biomass estimate for *actively* metabolizing microbes in communication with the atmosphere. There are other possible scenarios for extant life on Mars (see Fig. 3). For example, there could be an indefinitely large community of dormant microbes. Bacterial endospore states are highly durable to withstanding inclement environmental conditions and very low energy fluxes. Cellular turnover times of $\sim 10^3$ years have been found in deep biospheres on Earth (Lomstein *et al.* 2012), but the energy costs required of these dormant cells is unclear (Hoehler and Jorgensen 2013). Nevertheless, it is unlikely that spore-formation is evolutionarily advantageous for long-term survival in environments subject to harsh and degrading conditions, such as on Mars, due to the deterioration of DNA over time if molecular repair does not meet or exceed molecular damage (Cockell *et al.* 2016; Johnson *et al.* 2007; Mckay 1997; Teodoro *et al.* 2018).

Additional biomass could also be concentrated in self-sustaining isolated pocket communities that are closed-off from the atmosphere via impermeable lithological units (e.g. terrestrial microbes in Lin *et al.* 2006). However, these smaller isolated pockets would be harder to detect even with advanced drilling. One could also conceive of rarer unusual metabolisms that are in communication with the atmosphere but live off the regolith without excreting detectable products into the atmosphere. But these ecosystems seem unlikely, given the abundance of free energy available in the atmosphere and the simplicity and primitive nature of the CO-reaction genes in terrestrial microbes.

Finally, the results from our diffusion calculations place additional restrictions on the abundance and distribution of life. We find that CO and $H_2$ can diffuse into the subsurface at much higher rates than our maximum calculated biological sinks in the upper 500 m. This suggests that life is not taking full advantage of this available free energy in the upper regolith as the photochemical energy far outweighs alternative subsurface sources (Jakosky and Shock 1998). Therefore, any life in the upper crust is not limited by the available free energy but by some other factors (e.g. lack of liquid water or temperature dependency). Of course, this is consistent with the calculated instability of liquid water in the upper regolith, likely making it uninhabitable (Clifford et al. 2010). Future missions could also measure the CO and $H_2$ fluxes at the surface of Mars to place tighter constraints on the availability of reactants into the subsurface.

It should also be noted that we consider these diffusion calculations to apply to the low to mid-latitudes as the high latitudes will have an ice-saturated cryosphere, which is not considered here. The lower latitudes will be desiccated due to the instability of ground ice with respect to the water vapor content of the atmosphere (Clifford *et al.* 2010; Dundas *et al.* 2014; Feldman *et al.* 2004). Our calculations are relevant to the low to mid-latitudes because subsurface aquifers could be stable there over geological timescales (Grimm *et al.* 2017; Grimm and Painter 2009). As subsurface life presumably requires a liquid groundwater table, which is typically estimated around ≥5 km depth (Clifford *et al.* 2010), the potential diffusive flux may be smaller than our calculated maximum biogenic surface sinks at these depths and further limit the maximum allowable biomass unless a minimum water activity level for life is maintained by a minimum $H_2O$ layer thickness in the pore space somehow. Additionally, we note our use of a 1D photochemical model assumes a globally-distributed biosphere, but if life exists it could be more spatially limited, so the biomass estimate would presumably be even lower.

## 5. CONCLUSIONS

The martian atmosphere has an untapped free energy source of ~136 J mol$^{-1}$ predominantly contributed by the $CO$-$O_2$ (~133 J mol$^{-1}$) and $H_2$-$O_2$ (~3 J mol$^{-1}$) redox pairs (Krissansen-Totton *et al.* 2016). This constant photochemically-produced disequilibrium is far larger than any other energy sources known to exist on Mars, such as internal heat. Given life's tendency to exploit chemical free energy, we show how these relatively high concentrations of CO and $H_2$ coexisting with high concentrations of $O_2$ can represent antibiosignatures and that any extant life on Mars is severely limited.

We quantify how much life could be taking advantage of this 'free lunch' by using a 1-D photochemical code and modeling biological activity as a downward surface flux of CO and ramping it up until the modeled atmosphere diverges from observations. We find maximum feasible downward biogenic sinks of $1.5\times10^8$ molecules cm$^{-2}$ s$^{-1}$ and $1.9\times10^8$ molecules cm$^{-2}$ s$^{-1}$ for CO and $H_2$ respectively that are robust to uncertainties in observations and tunable model parameters. Using very conservative estimates on the minimal power requirements for microbial life and metabolic efficiencies, this equates to an upper limit of approximately $10^{27}$ cells or 1 million blue whales worth of metabolizing biomass that could be living off this atmospheric energy. This biomass estimate is highly dependent upon choice of cellular basal power requirements and typical microbes have higher requirements, thus any life is likely to be much smaller in extent. Diffusion calculations also imply that this biomass limit applies to life actively metabolizing within a few kilometers of the surface with the overall desiccation of the upper crust further limiting the possible biomass.

These results imply that any extant life on Mars is extremely limited in scope. CO and $H_2$ metabolisms are simple and phylogenetically widespread on Earth so that any microbial-like life on Mars should evolve to exploit this energy. Additionally, more plausible assumptions on metabolic efficiencies, combined with recognizing the maximum gas diffusion limit into the crust realistically only reach small pockets of liquid water, imply that any biomass may be orders of magnitude smaller than presented here. Thus, what we present here is a robust upper limit on extant actively metabolizing martian life in contact with the atmosphere.

Additionally, the concept of using sufficient atmospheric CO and $H_2$ concentrations as an antibiosignature could be applied to future research in exoplanet characterization. While the possibility for biogenic gases in exoplanet atmospheres as possible biosignatures has garnered much attention (e.g. Schwieterman *et al.* 2017; Seager and Bains 2015), the possibility of anti-biosignatures has received less (Catling *et al.* 2018; Wang *et al.* 2016). Where sufficient knowledge is known on the atmospheric composition, the presence of high concentrations of CO or $H_2$ in thermodynamic disequilibrium would allow for a deduction of an absence of a productive biosphere.

## ACKNOWLEDGEMENTS

We thank Drew Gorman-Lewis, Adam Stevens, and Kevin Zahnle for helpful discussions. We also thank Jim Kasting, Tullis Onstott, and an anonymous reviewer whose comments greatly improved the rigor and clarity of the manuscript. This work was supported by NASA Astrobiology Institute's Virtual Planetary Laboratory, grant NNA13AA93A, and by NASA Exobiology Program grant NNX15AL23G awarded to DCC. JKT is supported by NASA Headquarters under the NASA Earth and Space Science Fellowship program, grant NNX15AR63H. The paper was completed while DCC was a Leverhulme Trust Visiting Professor at the University of Cambridge, UK.

# **Tables**

**Table 1 - Mars' Atmosphere:** Modern martian atmospheric composition and uncertainties for modeled fluxes. Here we use the slightly higher background value of Webster *et al.* (2015) rather than that from Webster *et al.* (2018) for a more conservative upper limit on biomass.

| Species | Observed Mixing Ratio | Uncertainty (1 σ) | Reference |
|---------|----------------------|-------------------|-----------|
| $O_2$ | $1.74 \times 10^{-3}$ | $6 \times 10^{-5}$ | Franz *et al.* (2017) |
| CO | $7.47 \times 10^{-4}$ | $2.6 \times 10^{-6}$ | Franz *et al.* (2017) |
| $H_2$ | $1.5 \times 10^{-5}$ | $5 \times 10^{-6}$ | Krasnopolsky and Feldman (2001) |
| $CH_4$ | $6.9 \times 10^{-10}$ | $1.3 \times 10^{-10}$ | Webster *et al.* (2015)[a] |
| $CO_2$ | 0.96 | $7 \times 10^{-3}$ | Mahaffy *et al.* (2013) |

[a]These measurements are of a mean atmospheric level of 0.69 ppb and are controversial (see Zahnle *et al.* 2011), thus we place an upper bound of $9.4 \times 10^{-10}$ representing a 2σ detection.

**Table 2 – Parameter Space:** Unknown tunable model parameters and their assumed ranges for Mars. Tuned values are for the fixed-parameter abiotic modern Mars model. Grid space refers to how many equally spaced values of the parameter were used in the optimization grid search within the plausible range for the optimized-parameter model. $v_{dep}$ for CO, $CO_2$, $H_2$, and $O_2$ are set to 0 cm s$^{-1}$. The tuned $v_{dep}$ parameter refers to that assumed for all reactive species including both $O_3$ and $H_2O_2$.

| Parameter | Plausible Range | Grid Space | Tuned Valued | Reference(s) |
|---|---|---|---|---|
| Mean surface temperature, $T_s$ (K) | $199 - 215$ | 5 | 210 | Haberle (2013) |
| Surface deposition velocity, $v_{dep}$ (cm s$^{-1}$) | $0.001 - 0.1$ | 3 | 0.012 | Zahnle *et al.* (2008) |
| Ionospheric flux, $\varphi_{NO,CO}$ (molecules cm$^{-2}$ s$^{-1}$) | $10^1$ - $10^8$ | 5 | $7.3 \times 10^2$ | Krasnopolsky (1993); Smith *et al.* (2014) |

**Table 3 – Results:** Maximum allowable downward flux for each metabolism and the equivalent biomass. Biomass estimates are based on basal power requirements of $3\times10^{-23}$ kJ s$^{-1}$ cell$^{-1}$ (LaRowe and Amend 2015; Lever *et al.* 2015) and an average cellular mass of $7\times10^{-17}$ kg (Bratbak and Dundas 1984; Hoehler and Jorgensen 2013). The column headed 'Break' describes which atmospheric species deviates from observed abundance first and whether it is higher or lower than the 2σ detection. The greatest biomass estimate model runs are highlighted in bold.

| # | Metabolism | Net Reaction | Fixed | | | | Optimized | | | |
|---|---|---|---|---|---|---|---|---|---|---|
| | | | Max Flux [mol. cm$^{-2}$ s$^{-1}$] | Break | Biomass [cells] | Biomass [kg] | Max Flux [mol. cm$^{-2}$ s$^{-1}$] | Break | Biomass [cells] | Biomass [kg] |
| 1 | Anaerobic CO-only | $CO + H_2O \rightarrow CO_2 + H_2$ | $8.4 \times 10^5$ | CO↓ | $1.6 \times 10^{24}$ | $1.1 \times 10^8$ | $1.4 \times 10^8$ | CO$_2$↓ | $2.6 \times 10^{26}$ | $1.8 \times 10^{10}$ |
| 2 | Methanogenesis | $CO_2 + 4H_2 \rightarrow CH_4 + 2H_2O$ | $1.4 \times 10^5$ | CH$_4$↑ | $2.6 \times 10^{23}$ | $1.8 \times 10^7$ | $1.9 \times 10^5$ | CH$_4$↑ | $3.5 \times 10^{23}$ | $2.5 \times 10^7$ |
| 3 | Anaerobic CO and methanogenesis | $4CO + 2H_2O \rightarrow 3CO_2 + CH_4$ | $1.4 \times 10^5$ | CH$_4$↑ | $5.2 \times 10^{23}$ | $3.7 \times 10^7$ | $1.6 \times 10^5$ | CH$_4$↑ | $5.9 \times 10^{23}$ | $4.2 \times 10^7$ |
| 4 | Aerobic CO-only | $2CO + O_2 \rightarrow 2CO_2$ | $2.6 \times 10^6$ | CO↓ | $5.3 \times 10^{25}$ | $3.7 \times 10^9$ | **$1.5 \times 10^8$** | **CO$_2$↓** | **$3.1 \times 10^{27}$** | **$2.2 \times 10^{11}$** |
| 5 | Methanogenesis and methanotrophy | $O_2 + 2H_2 \rightarrow 2H_2O$ | $4.4 \times 10^6$ | CO↑ | $7.3 \times 10^{25}$ | $5.1 \times 10^9$ | **$1.9 \times 10^8$** | **H$_2$↓** | **$3.2 \times 10^{27}$** | **$2.2 \times 10^{11}$** |

# Figures

**Fig. 1:** Modeled atmospheres for Mets. 1-3 as biogenic sinks are incrementally ramped up. Left panels are for the fixed parameter model and the right panels for the optimized parameter version (see text). Shaded regions represent a 2σ uncertainty for the mixing ratios of CO, $O_2$, $H_2$, $CH_4$, and $CO_2$. Vertical dotted lines indicate where the model diverges from observations (max biological sink). $CO_2$ concentrations do not vary for Met. 2 within or below this biogenic sink range.

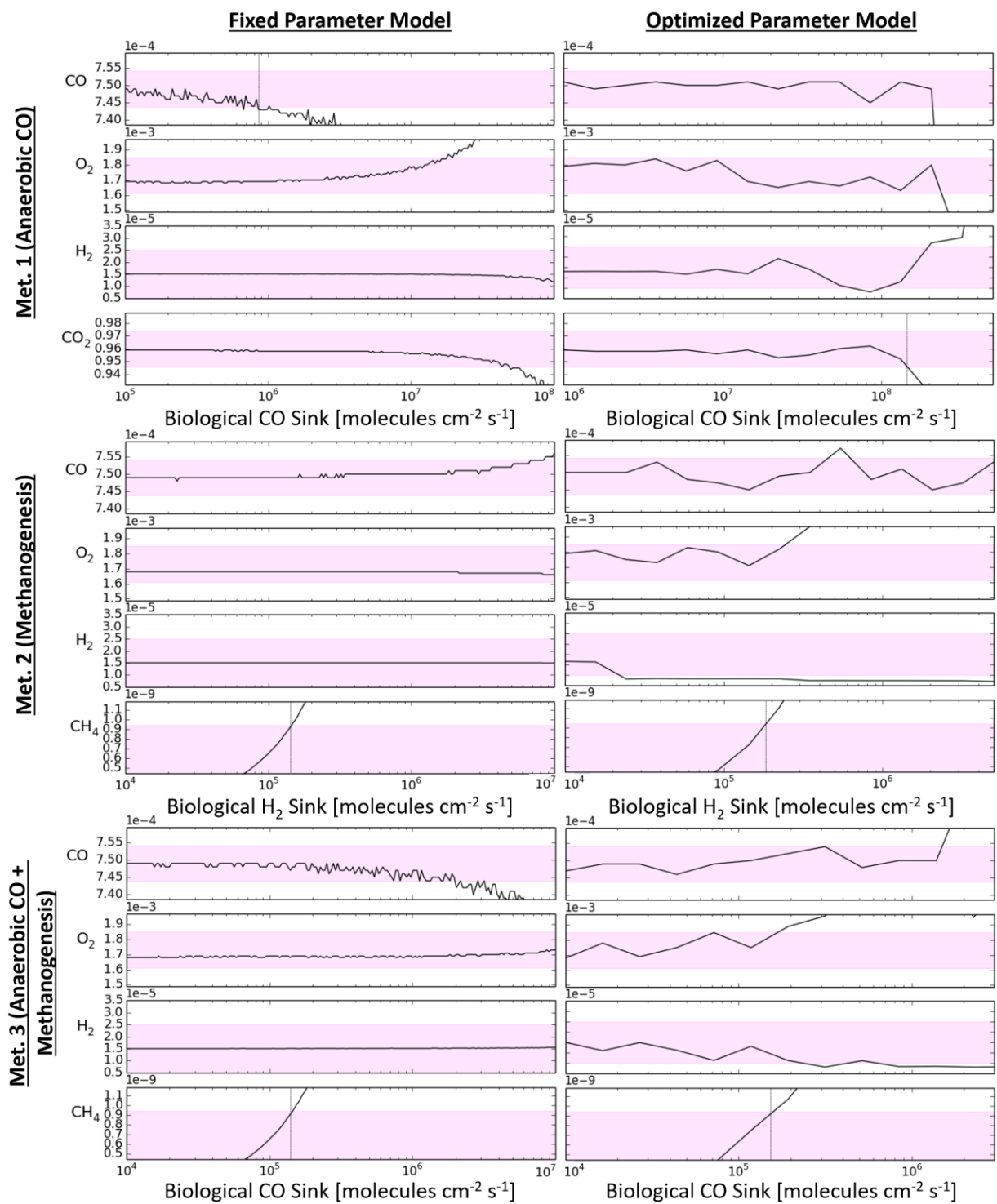

**Fig. 2**: Modeled atmosphere for Mets. 4 and 5, which produce the largest maximum biomass, as biogenic sinks are incrementally ramped up. Left panels are for the fixed parameter model and the right panels for the optimized parameter version (see text). Shaded regions represent a $2\sigma$ uncertainty for the mixing ratios of CO, $O_2$, $H_2$, $CH_4$, and $CO_2$. Vertical dotted lines indicate where the model diverges from observations (max biological sink). $CO_2$ concentrations do not vary for Met. 5 within or below this biogenic sink range.

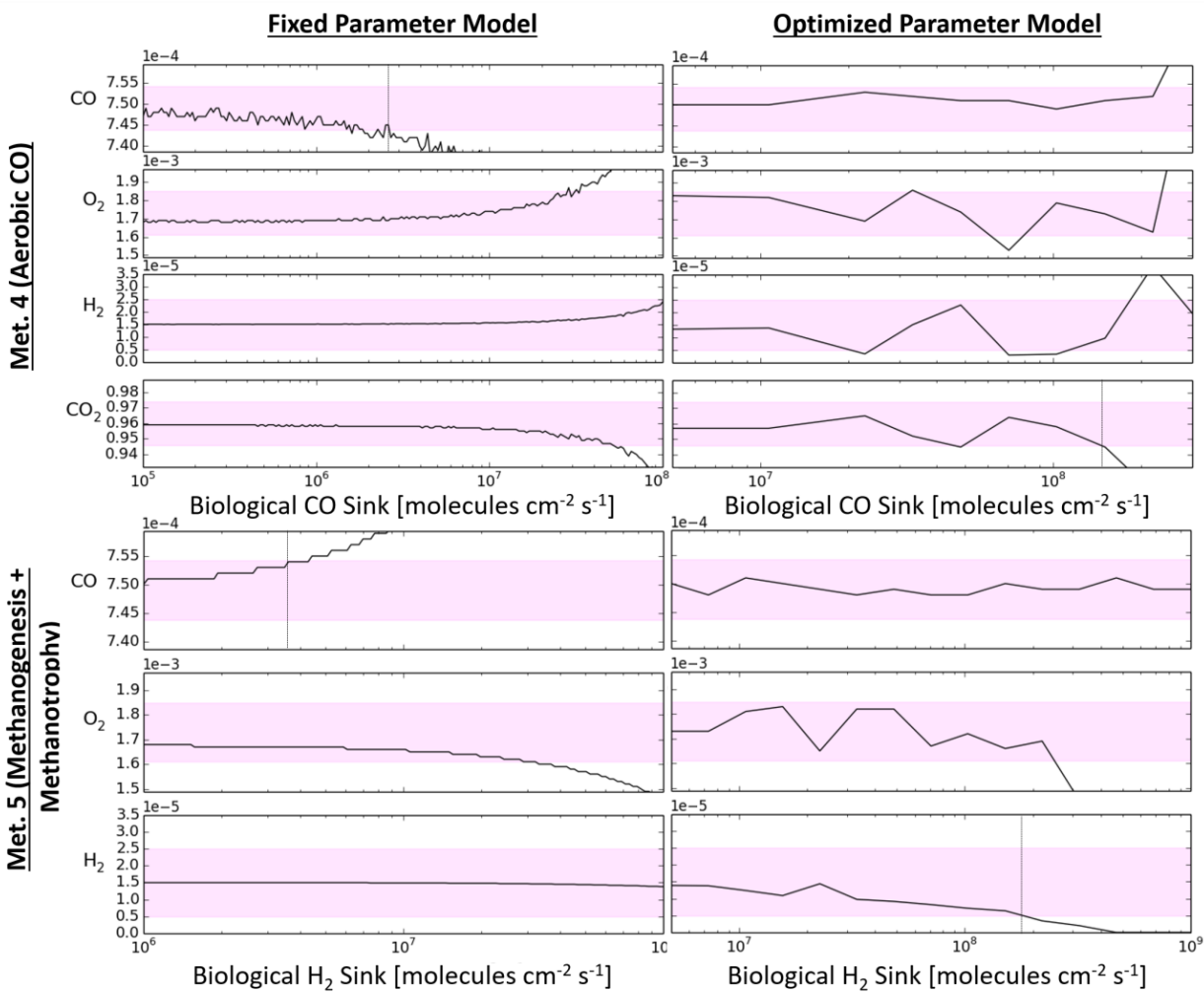

**Fig. 3:** Monte Carlo simulations calculating diffusive fluxes for CO and $H_2$ as a function of biotic layer depth. The resulting probability density is shown via colored bins. Solid black trendlines indicate median flux values while vertical lines show maximum allowable biogenic sinks for each metabolism from our photochemical model calculations. The interception of these vertical lines with the median diffusion flux shows the maximum depth to which subsurface life could be exploiting atmospheric free energy. Below these depths, microbes would be limited by the downward diffusion of atmospheric gases through the regolith rather than the supply of reactants from photochemical reactions.

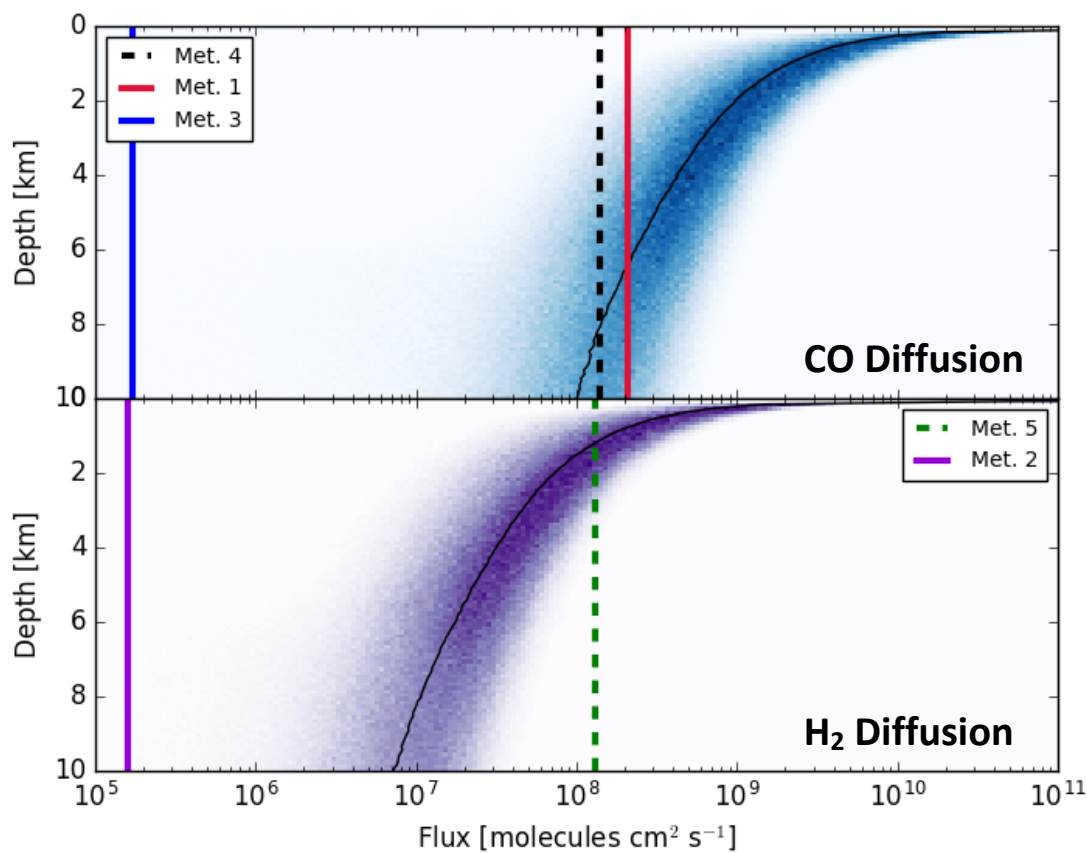

**Fig. 4:** Potential sites for life on Mars. This study estimates the maximum number of actively metabolizing microbes that are in communication with the atmosphere (1). There could be actively metabolizing microbes in small communities that are sealed off from the atmosphere (2) or are neither taking advantage of the available atmospheric free energy nor producing detectable byproducts (3). The number of dormant microbes (e.g. endospores) (4) could be indefinitely large, but would not be evolutionarily advantageous for long-term survival on Mars. Downward arrows indicate fluxes of available free-energy reactants and upward arrows indicate flux of metabolized products into the atmosphere. All are assumed to be in contact with some form of liquid water (e.g. rare briny fluid pockets or a deep aquifer).

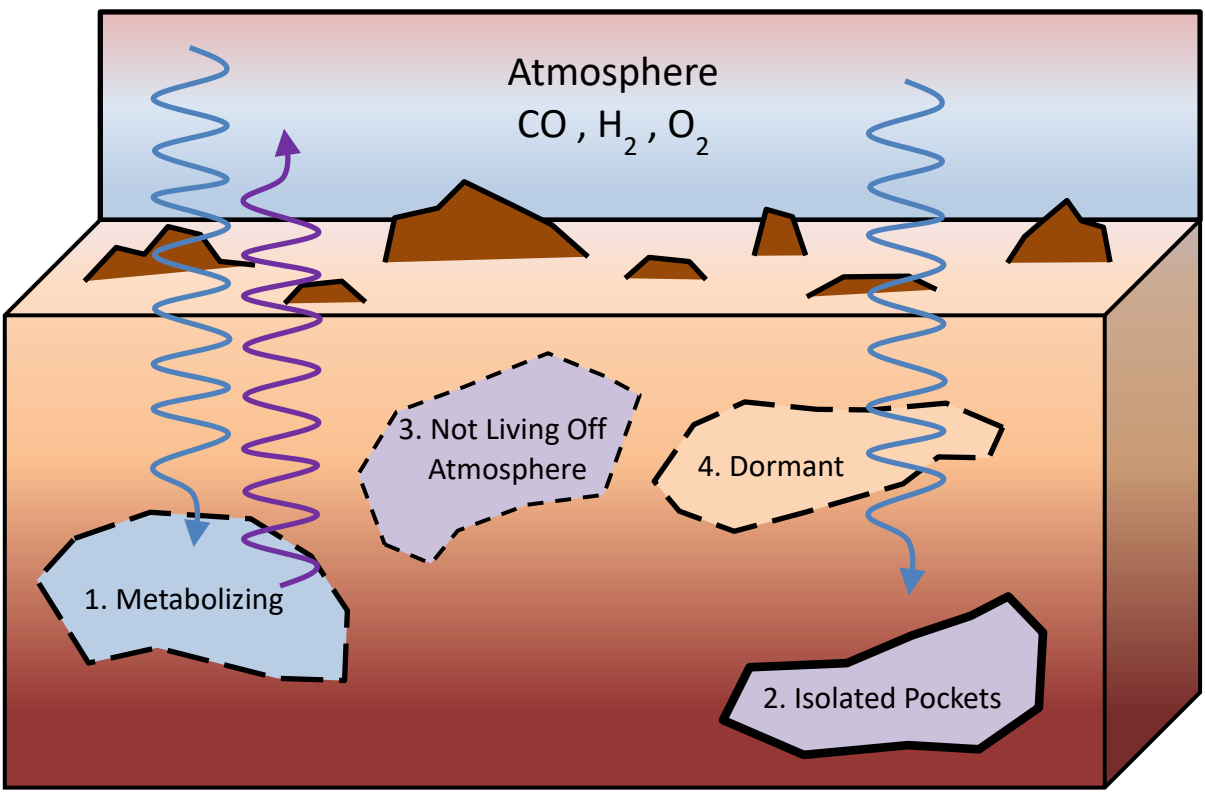